
%
%
%
\input harvmac
\vsize=7.5in
\hsize=5in
\pageno=0
\tolerance=10000
\def\uplrarrow#1{\raise1.5ex\hbox{$\leftrightarrow$}\mkern-16.5mu #1}
\def\bx#1#2{\vcenter{\hrule \hbox{\vrule height #2in \kern #1\vrule}\hrule}}

\def\tr{\,{\hbox{tr}}\,}
\def\Tr{\,{\hbox{Tr}}\,}

\def\squiggle#1{\lower1.5ex\hbox{$\sim$}\mkern-14mu #1}

\def\thru#1{\mathrel{\mathop{#1\!\!\!\!/}}}

\def\same#1#2{\mathrel{\mathop{#1\!\!\!\!\!\!\!\!#2}}}

\def\narrower{\advance\leftskip by\parindent \advance\rightskip by\parindent}

\def\mbox#1#2{\vcenter{\hrule width#1in\hbox{\vrule height#2in
   \hskip#1in\vrule height#2in}\hrule width#1in}}
\def\eqsquare #1:#2:{\vcenter{\hrule width#1\hbox{\vrule height#2
   \hskip#1\vrule height#2}\hrule width#1}}
\def\inbox#1#2#3{\vcenter to #2in{\vfil\hbox to #1in{$$\hfil#3\hfil$$}\vfil}}
\def\strutdepth{\dp\strutbox}
\def\marbul{\strut\vadjust{\kern-\strutdepth\specialbul}}
\def\specialbul{\vtop to \strutdepth{
    \baselineskip\strutdepth\vss\llap{$\bullet$\qquad}\null}}
\def\Bcomma{\lower6pt\hbox{$,$}}    
\def\bcomma{\lower3pt\hbox{$,$}}    

\def\updots{\mathinner{\mskip 1mu\raise 1pt\hbox{.}
    \mskip 2mu\raise 4pt\hbox{.}\mskip 2mu
    \raise 7pt\vbox{\kern 7pt\hbox{.}}\mskip 1mu}}

\def\pmb#1{\setbox0=\hbox{#1}%
     \kern-.025em\copy0\kern-\wd0
     \kern.05em\copy0\kern-\wd0
     \kern-.025em\raise.0433em\box0}

\def\1{\;1\!\!\!\! 1\;}
\def\eg{{\it e.g.}}
\def\ie{{\it i.e.}}

\def\m@th{\mathsurround=0pt}
\def\upsquarefill{$\m@th\bracelu\leaders\vrule\hfill\braceru$}
\def\ope#1{\mathop{\vtop{\ialign{##\crcr
     $\hfil\displaystyle{#1}\hfil$\crcr\noalign{\kern3pt\nointerlineskip}
     \kern4pt\upsquarefill\kern4pt\crcr\noalign{\kern3pt}}}}\limits}
\def\Int#1#2{\int\!d^{#1}{#2}\,}
\def\lsim{\mathrel{\rlap{\lower4pt\hbox{\hskip1pt$\sim$}}
    \raise1pt\hbox{$<$}}}         
\def\gsim{\mathrel{\rlap{\lower4pt\hbox{\hskip1pt$\sim$}}
    \raise1pt\hbox{$>$}}}         

\hfuzz=5pt
\baselineskip 12pt plus 2pt minus 2pt
\hfill hep-th/9303113

\hfill DFTT 10/93

\hfill March 1993
\centerline{\bf ANOMALY--INDUCED MAGNETIC SCREENING}
\centerline{\bf IN 2+1 DIMENSIONAL QED AT FINITE DENSITY}
\vskip 36pt\centerline{Stefano
Forte}
\vskip 12pt
\centerline{\it I.N.F.N., Sezione di Torino}
\centerline{\it via P.~Giuria 1, I-10125 Torino, Italy}
\vskip 1.7in

{\narrower\baselineskip 12pt
\centerline{\bf ABSTRACT}
\noindent
We show that in $2+1$ dimensional Quantum Electrodynamics
an external magnetic field applied to a finite density
of massless fermions is screened, due to a
$2+1$-dimensional realization of the  underlying
$2$-dimensional axial anomaly of the space components of the
electric current. This is shown to imply screening of the magnetic
field, \ie, the Meissner effect. We discuss the physical
implications of this result.}

\vskip 1.in
\centerline{Submitted to: {\it Physical Review Letters}}
\vfill
\eject
Quantum Electrodynamics in $2+1$ dimension has attracted recently considerable
theoretical attention, due to both theoretical and experimental motivation from
condensed matter systems: on the one hand, several startling features of planar
quantum mechanics, such as fractional statistics\nref\rmp{For a review
see S.~Forte, {\it Rev. Mod. Phys} {\bf 64}, 193 (1992)}$^{(\xref\rmp)}$
and Chern-Simons field theory\nref\rom{For a review
see R.~Jackiw {\it Nucl. Phys. (Proc. Suppl.)} {\bf 18A}, 107 (1991)}$^{(
\xref\rom)}$ have started revealing themselves; on the other hand both the
Quantum Hall Effect and High-T$_c$ superconductivity are crucially related to
the effectively two-dimensional nature of the systems where they have been
discovered. Indeed, the possibility of excitations carrying arbitrary
statistics, effectively described by coupling of the charged excitations to
a Chern-Simons field theory, is at the basis of the most succesful theory of
the Quantum Hall effect,\nref\pran{See
``The Quantum Hall Effect'', R.~E.~Prange and S.~M.~Girvin, Eds. (Springer,
Berlin, 1987)}$^{(\xref\pran)}$ the Laughlin theory\nref\laugh{See
the contribution of R.~B.~Laughlin
in Ref.\pran}$^{(\xref\laugh)}$, and is perhaps at the root of high T$_c$
superconductivity\nref\wilcz{See F.~Wilczek, ed. ``Fractional Statistics and
Anyon Superconductivity'' (World Scientific, Singapore, 1990)}$^{(
\xref\wilcz)}$.

The appearance of superconducting behavior in $2+1$ dimensional systems with
fractional statistics seems to be a rather robust feature of any theory of
fermions coupled to Chern-Simons fields\nref\lykk{J.~D.~Lykken, J.~Sonnenschein
and N.~Weiss, {\it Phys. Rev.} {\bf D42} (rap. comm.), 2161 (1990)
}$^{(\xref\lykk)}$, at least at zero temperature. Because a Chern-Simons
interaction is always generated dynamically by quantum fluctuations in the
effective action of $2+1$-dimensional fermions coupled to  external gauge
potentials\nref\redl{A.~N.~Redlich, {\it Phys. Rev. Lett.} {\bf 52}, 1983
(1984)}$^{(\xref\redl)}$, unless forbidden by symmetry, it is quite conceivable
that superconductivity induced by the topological structure of the theory which
is  associated to fractional statistics
should be a more general feature of
$2+1$-dimensional electrodynamics.

Here, we shall show that this is indeed true in a well-defined sense: in
$2+1$-dimensional electrodynamics with a chemical potential, \ie, with
finite fermion density, an analogue of the London equation always holds; this
(for magnetic fields below a critical value)
leads to the Meissner effect, \ie, screening of the magnetic field,
which is a hallmark of superconductivity.

Rather
than analyzing the effective action of fermions coupled to the gauge
potentials, and discussing possible relations of our result to the usual
approaches based on an effective Chern-Simons coupling and fractional
statistics, we shall establish our result by direct computation of the
vacuum expectation value of the electric current and its curl. We shall show
that the the electric current, \ie, the space components of the
three-dimensional covariant fermionic current of the theory, satisfies,
to first order in the magnetic field, a
$2+1$-dimensional analogue of the anomaly equation in two dimensions.
This gives to its curl an
anomalous, non-vanishing vacuum expectation value,
proportional to the magnetic field itself, with a coefficient that depends
on the parity-breaking parameter, \ie, on the chemical potential.
 Thus, quantum fluctuations induce anomalous
vacuum currents which screen the applied magnetic field, with a screening
length depending on the chemical potential. We shall establish this result for
time-independent gauge fields, as well as for time dependent ones at the
leading order in the adiabatic approximation, in the low temperature
limit.
We shall also discuss
briefly the physical relevance of our
results.

We study 2+1 dimensional massless electrodynamics with a chemical potential
term, defined by the Lagrangian
\eqn\lag{L=\Int 3 x  \left[-\tr{1\over 4 e^2} F_{\mu\nu} F^{\mu\nu}+
\psibar\left( \gamma^\mu\left(i \partial_\mu -  A_\mu\right)-\mu
\gamma^0\right)\psi
\right],}
where the coupling $e^2$ has the dimensions of a mass.
We shall work throughout in the $A_0=0$ gauge.
 Our starting point is the observation that the spatial components of the
fermion current in 2+1-dimensional Minkowski space-time are closely related
to the
axial current of a 2-dimensional Euclidean theory. Indeed, the fermion current
of the theory \lag\ is
\eqn\cum{j^\mu=\psibar\gamma^\mu\psi,}
where we can
represent the 2+1-dimensional Dirac matrices in terms of Pauli matrices as
$\gamma^i=i\sigma^i$, ($i=1,2$), $\gamma^0=\sigma^3$.
Consider now the usual Dirac $\alpha$ matrices, $\alpha^i=\gamma^0\gamma^i$,
which
satisfy a two-dimensional Euclidean Clifford algebra $\{\alpha^i,
\alpha^j\}=\delta^{ij}$, and may thus  be used to construct a Euclidean
Dirac operator
\eqn\euc{\eqalign{i\thru D&=(i\del_i-A_i) \alpha^i-\mu\cr
&\equiv -H^{(\mu)}=-\left(H+\mu\right).\cr}}
The Dirac operator \euc\ is readily recognized to coincide with (minus)
the Hamiltonian
$H^{(\mu)}$ of the given theory \lag; notice that the chemical potential term
is
formally identical to a two-dimensional (imaginary) mass term.
Recalling that in a two-dimensional Euclidean theory
$\gamma_5=i\gamma^1_{(2)}\gamma^2_{(2)}$, it follows that
the axial current associated to the Dirac operator \euc\ is
\eqn\cue
{\eqalign{{j^i_5}_{(2)}&=i\psi^\dagger\gamma^i_{(2)}\gamma_5\psi\cr
&=i\psi^\dagger\alpha_i\sigma_3\psi ,\cr}}
which is related to the spatial
components of the $2+1$-dimensional Minkowski current \cum\
by
\eqn\currel
{{j^i_5}_{(2)}=\epsilon^{ij}j^i,}
where $\epsilon^{ij}$ is the completely antisymmetric tensor in two dimension
normalized as $\epsilon^{12}=1$.

Now, the current \cue\  satisfies an axial anomaly equation\nref\anrev{See
R.~Jackiw, in S.~B.~Treiman, R.~Jackiw, B.~Zumino and E.~Witten, ``Current
Algebra and Anomalies'' (World Scientific, Singapore, 1985)}$^{(\xref\anrev)}$
\eqn\aan{\del_i{j^i_5}_{(2)}=-{1\over\pi} B(x)}
where $B(x)=\epsilon_{ij}\partial^i A^j(\vec x)$.
We are thus led to conjecture
that for time-independent fields, and more generally at the
leading order in the adiabatic approximation, the spatial part of the
current $j^\mu_{(3)}(t,\vec x)$ at fixed time $t$ should satisfy a similar
equation. This Eq. would then  have the structure of a London equation.
A minute's reflection reveals that the conjecture cannot be literally true, for
the two-dimensional current \cue\ and the three-dimensional one \cum\ do not
have the same dimensions, thus the space components of the latter cannot
satisfy Eq.\aan\ as is. Also, an equation of the form \aan\
for the space components of the current signals breaking of
parity in the 2+1 dimensional theory (just as the presence of a Chern-Simons
term in the action) and can therefore be induced by radiative correction only
in the presence of a parity-breaking parameter. As we shall see, the
presence of a chemical potential will provide both the parity breaking, and the
required dimensionful parameter.

We shall now proceed to establish our main result, namely, prove the conjecture
that the space components of the current \cum\ satisfy a two-dimensional
anomaly equation of the form \aan.
To this purpose,
we consider the
$\zeta$-function regularized\nref\niese{See A.~J.~Niemi and G.~W.
{}~Semenoff {\it Phys. Rep.} {\bf 135}, 99 (1986); P.~B.~Gilkey ``Invariance
Theory, the Heat Equation, and the Atiyah-Singer Index Theorem'' (Publish or
Perish, Wilmington, De, 1984)}\nref\hawk{S.~W.~Hawking,
{\it Comm. Math. Phys.} {\bf 55}, 133 (1977)}$^{(\xref\niese,\xref\hawk)}$
expression for the vacuum expectation value of the current\nref\zetan{
Notice that no subtraction is needed in that the analytic continuation of the
$\zeta$-function provides automatically both the required regularization and
renormalization. It is easy to generalize to the current thus defined the
standard derivation of the anomaly equation based upon Pauli-Villars
regularization.}$^{(\xref\zetan)}$ Eq.\cum:
\eqn\reg
{\langle j^\mu(x)\rangle=
\lim_{s\to1}\tr \left[ \gamma^\mu R(x,x;s)\right],}
where $R(x,y;s)$ is the $\zeta$-function regularization
of the resolvent
of the Dirac operator which appears in \lag, \ie,
\eqn\res
{\eqalign{&i\thru D_3 \lim_{s\to 1} R(x,y,;s)=\delta^{(3)}(x-y)\cr
&\quad i\thru D_3=\left(i{\partial\over\partial t}-H^{(\mu)}\right),\cr}}
where $H^{(\mu)}$ is given by Eq.\euc.
Explicitly we have
\eqn\resdef
{R(x,y,;s)=\same\int\sum_k{\hat\psi_k(x)\hat\psi^\dagger(y)
\over\lambda_k^s}=\Tr \langle x| {1\over \left[\thru D\right]^s} |
 y\rangle,}
where $i\thru D_3\hat\psi_k=\lambda_k\hat\psi_k$,
and the sum, which runs over the full 2+1-dimensional
spectrum, has been indicated formally as a pointwise trace in the last step,
which in the $x\to y$ limit we shall henceforth
denote with $\Tr$, for short.

We may now compute the curl of the regulated vacuum expectation value of the
current \reg:
\eqn\curres
{\langle\epsilon^{ij}\partial_i J^j\rangle=\lim_{s\to1}\Tr\left[
\alpha^j\epsilon^{ij}\partial_i{1\over\left[ i {\partial\over\partial
t}-\left(H+\mu\right)\right]^s}\right].}
In order to determine the functional trace in Eq.\curres\ we diagonalize the
Dirac operator in the adiabatic approximation
in terms of the eigenfunctions of the Hamiltonian $H$, by compactifying time
over a period $T$, which we shall let eventually to infinity in the
zero-temperature limit, and introducing antiperiodic boundary conditions over
$T$.\nref\adiab{
See A.~J.~Niemi and G.~W.~Semenoff, {\it Phys. Rev. Lett.} {\bf 55},
927 (1985); S.~Forte and
P.~Sodano {\it Nuovo Cim.} {\bf D11}, 321 (1989) and ref. therein.}$^{
(\xref\adiab)}$ The eigenfunctions and eigenvalues of $i\thru D_3$ \res\
are then given by
\eqn\spec
{\eqalign{&i\thru D_3\hat\psi_{nk}=\lambda_{nk}\hat\psi_{nk}\cr
&\qquad\lambda_{nk}={(2k+1)\pi\over T}- \left(\alpha_n+\mu\right)\cr
&\qquad \hat\psi_{nk}(\vec x,t)={1\over T}e^{i\lambda_{nk} {t\over T}}
\psi_n(\vec x)\cr}}
in terms of the so-called Floquet indices$^{(\xref\adiab)}$
$\alpha_n$ of
the instantaneous eigenfunctions and eigenvalues of the Hamiltonian
\euc:
\eqn\spech
{\eqalign{& H(\vec x; t)\psi_n(\vec x, t)=E_n(t)\psi_n(\vec x, t)\cr
&\qquad \alpha_n={1\over T} \int_0^T \!dt\, E_n(t).
\cr}}
Of course this is the exact spectrum in the case of  time-independent gauge
fields, when $\alpha_n=E_n$.

Substituting Eq.s\spec,\spech\ in the expression \curres\ for the curl of the
current we get
\eqn\restrac{\langle\epsilon^{ij}\partial_i J^j\rangle=
-\lim_{s\to1} \partial_i\same\int\sum_{n}\sum_{k=-\infty}^{\infty}\tr\, i
\alpha^i\gamma_5
{{1\over T}\psi_{n}(x)\psi_{n}^\dagger(x)\over\left[{(2k+1)\pi\over
T}+\alpha_n+\mu\right]^s},}
whence, in the particular case of time-independent fields,
\eqn\resfin{\eqalign{\langle\epsilon^{ij}\partial_i J^j\rangle&=
\lim_{s\to1} \same\int\sum_{k,n} 2 E_n
\tr {\psi_{n}^\dagger\gamma_5(x)\psi_{n}(x)\over\left[{(2k+1)\pi\over
T}+E_n+\mu\right]^s}\cr
&= 2 \lim_{s\to1} \Tr\, \gamma_5 H \left( D+\mu + H\right)^{-s}
,\cr}}
where in the last step $D=-i{\partial\over\partial t}$, $\gamma_5=\sigma_3$,
and we have reverted to
the functional trace notation which now indicates both summation over the
eigenfunctions ${(2k+1)\pi\over T}$ of $D$ and summation and integration over
the full spectrum of the Hamiltonian $H$, Eq.\euc. It is easy to see that
Eq.\resfin\ holds in general in the time-dependent case (at leading adiabatic
order) for the time-average of the curl, provided the eigenvalues $E_n$ are
replaced everywhere by the Floquet indices  $\alpha_n$ \spech.

Once the trace over eigenfrequencies of the Dirac operator $i\thru D_3$ has
been reduced to the spectra of the operators $D$ and $H$ according to Eq.
\resfin, we may proceed to compute the latter.
We shall make first use of the fact that
both the operators $H$ and $D$ have an
antisymmetric spectrum, as proven by the existence of an operator
which anticommutes with
each of them, namely $\gamma_5$ and the time-reversal operator
$\cal T$, respectively:
$\{\gamma_5, H\}=\{{\cal T}, D\}=0$.
Using  the antisymmetry of $D$ first, we get
\eqn\rep
{\langle\epsilon^{ij}\partial_i J^j\rangle=2\lim_{s\to1}
\Tr\,\gamma_5 H\left(\mu+H\right)\left[\left(\mu+H\right)^2-
D^2\right]^{-s}.
}

We may now separate the functional traces over $D$ and $H$
by taking a Mellin transform, whereby Eq.\rep\
becomes
\eqn\lap
{\langle\epsilon^{ij}\partial_i J^j\rangle=
-2\lim_{s\to1}\int\!{dt\over\Gamma(s)}\,t^{s-1}\Tr_D\left[
e^{- t D^2}\right]\Tr_H \left[\gamma_5\left(\mu H+H^2\right)
e^{t (\mu+H)^2}\right]
,}
where $\Tr_D$ and $\Tr_H$ denote the traces of the respective operators, which
are to be evaluated, as usual, by Wick-rotating to Euclidean space and letting
$T\to\infty$, in order to enforce the correct (Feynman) boundary conditions.
Upon Wick rotation the spectrum of $H$, as well as the chemical potential
$\mu$ become purely imaginary, thereby ensuring convergence of the $t$
integration, and
allowing the Mellin transform that has
lead  to Eq.\lap, at least if $s$ is kept large enough.

The trace over $D$ may now be computed explicitly in the large $T$ limit:
\eqn\trd
{\eqalign{\lim_{T\to\infty}\Tr_D\left[e^{- t D^2}\right]=
&{1\over 2\pi}\int_{-\infty}^{\infty}\!dk
e^{-t k^2}\cr
&={1\over2\sqrt{\pi t}}\cr}}
Using this result in Eq.\lap\ and inverting the Mellin transform we get
\eqn\trh
{\langle\epsilon^{ij}\partial_i J^j\rangle=
-\lim_{s\to1}\Tr_H \gamma_5 \left(\mu H+H^2\right)
\left[\left|\mu+H\right|\right]^{-s}.}
Before computing the trace over the spectrum of the Hamiltonian we may further
simplify Eq.\trh, using the antisymmetry of the spectrum of $H$:
\eqn\trhsq
{\langle\epsilon^{ij}\partial_i J^j\rangle=
\lim_{s\to 0}\widetilde{\Tr}\, \gamma_5 \left|H\right|
\left[H^2-\mu^2\right]^{-s},}
where $\widetilde{\Tr}$ denotes the trace over all eigenfrequencies
of $H$ such that the respective eigenvalue $\lambda$ satisfies
$\lambda > \mu$ (\ie, such that the denominator in Eq.\trhsq\ is positive
definite).

The functional trace may now be computed explicitly, observing that
in position space its
diagonal matrix element is given by
\eqn\trdiag
{\eqalign{\lim_{s\to 0}\langle x|\left[
\widetilde{\Tr} \gamma_5 \left|H\right|
\left[H^2-\mu^2
\right]^{-s}\right]&|x\rangle=
\cr&\!\!\!\!\!\!\!\!\!\!\!\!\!=\lim_{s\to 0}\tr
\int_{|\mu|}^\infty\!{dk\over2\pi}\, k\sqrt{k^2 + \gamma_5 B(x)}
\left[k^2+\gamma_5 B(x) -\mu^2\right]^{-s}\cr
&\!\!\!\!\!\!\!\!\!\!\!\!\!=-{1\over2\pi}|\mu| B(x)+O\left(\left[
{B\over \mu^2}\right]^2\right),\cr}}
where $B(x)$ is as in Eq.\aan, and
as in the standard functional
derivation of the anomaly,\nref\bro{L.~S.~Brown, R.~D.~Carlitz and C.~Lee,
{\it Phys. Rev.} {\bf D16}, 417 (1977)}$^{(\xref\bro)}$ we used the fact that
\eqn\hsq
{\eqalign{&H^2=D_iD_i+\gamma_5 B(x)\cr
& D_i=-i\partial_i +A_i
\cr}}
while, since the trace as $s\to 0$ is dominated by its large-momentum behavior,
the spectrum of $D_i$ may be approximated by that of
$\partial_i$.\nref\foota{The result \trdiag\ is obtained
expanding all functions of $H^2 -\mu^2$ in powers of
${B-\mu^2\over D^2}$.
Expanding in powers of ${B\over D^2-\mu^2}$ instead would produce a spurious
infrared divergence which leads to a vanishing result for Eq.\trdiag
unless it is regulated by an infrared mass, in which case the result \trdiag\
is reproduced. It is
easy to verify that the same infrared divergence would also lead to vanishing
of the axial anomaly.
}$^{(
\xref\foota)}$

 Using the expression of the trace \trdiag\ in Eq.\trhsq\
we get finally
\eqn\vict
{\langle\epsilon^{ij}\partial_i J^j(x)\rangle=
-{1\over2\pi}|\mu|B(x)+O\left(\left[
{B\over \mu^2}\right]^2\right)}
which is our main result, and proves our conjecture:
the vacuum-expectation value of the curl of the current is
anomalous, and, to first order, proportional to the magnetic field.
Notice that this
a purely quantum effect, in that it pertains a vacuum
expectation value of the electric current, which is generated by quantum
fluctuations.

If we neglect higher order corrections  under the assumption that the magnetic
field is not too  strong, Eq.\vict\
is immediately recognized as the London equation, which, together with the
2+1-dimensional Maxwell equation
\eqn\maxw
{J^i={1\over e^2}\epsilon^{ij}\partial_j B(x)}
leads to the
screening equation in the plane
\eqn\scree
{\Delta B(x)={e^2|\mu|\over 2\pi} B(x),}
\ie, to the Meissner effect with screening length
\eqn\lscr
{\lambda=\sqrt{2\pi\over e^2|\mu|}.}

Let us now discuss the physical implications
of this result.
If
a layered physical system may be considered as
effectively two-dimensional, and we take the Lagrangian \lag\ as the
effective theory of its charged excitations,
with the parameter $e^2$ related to the typical distance between the
planes which build up the three-dimensional bulk of matter,
then a constant magnetic field applied to that material
is  screened, provided in the given region $|B|< |\mu|^2$. For example,
for high-$T_c$ layered superconductors typical values
are\nref\chak{See \eg, Y.~Hosotani and S.~Chakravarty, {\it Phys. Rev.}
{\bf B 42}, 342 (1990)}$^{(\xref\chak)}$
$e^2\sim 20$ eV. With a chemical potential $\mu$ of order of a few eV the
screening length is $\lambda\sim 10^3$ \AA, as in ordinary superconductors.
When the magnetic flux density exceeds $\mu^2$, the approximation
leading from \vict\ to \scree\ breaks down; if we take this as an estimate
of the critical flux density we get $B_c\sim (10^3\> {\rm \AA})^{-2}$, \ie,
$B_c\sim 10^3$ gauss, as in usual superconductors.

Even though the present discussion has been given in the low-temperature
limit, this assumption has been used only in the computation of the
trace over Matsubara frequencies, Eq.s \resfin-\lap, which has been
performed in the limit $T\to\infty$. It actually turns out that the first
order correction to these results, which is of order $\mu\over T$, vanishes,
thus suggesting that our results persist up to fairly high temperature
(notice that $\mu
\sim 10^4~{}^\circ K$). Also, even though the chemical potential has played a
crucial role in our discussion, one may notice that beyond the leading
adiabatic approximation (to which we have confined ourselves) the
spectrum \spec\ also receives$^{(\xref\adiab)}$ a
Berry phase contribution which enters
Eq.\spec\ in a way formally identical to the chemical potential. This suggests
that  by retaining such
contributions our result may be generalized to zero density.

Finally, it is interesting to observe that the induced anomalous currents which
screen the magnetic field  according to Eq.\vict\ are transverse,
as those which would be generated by a Chern-Simons term in the effective
action of the theory, a characteristic feature of the Hall effect.
This suggests
the possibility of applying the present results to the Hall system.
These and related problems are currently under investigation.

In sum, we have proven that the electric current of fermions in $2+1$
dimensions has an anomalous curl induced by quantum fluctuations, and related
to the chiral anomaly of an underlying two-dimensional Euclidean theory.
This curl obeys the Meissner equation and thus it induces
screening of the
magnetic field.
This  provides a
firmer footing to the observation that topological and superconducting behavior
seem to be rather robust features of gauge theories in $2+1$ dimensions.
It would be interesting to try to apply the present considerations to the
realistic lattice systems which display high-T$_c$ superconductivity:
perhaps this approach may provide some insight in the
sought-for novel mechanism which is at the origin of these phenomena.

{\bf Acknowledgement:} This works is based on an idea developed in discussions
with P.~Sodano, and reported in Ref.\ref\prelim{S.~Forte and P.~Sodano, Torino
University Report DFTT 35/92 (unpublished)}. I thank P.~Sodano for
collaboration in the early stages of this work, and a critical reading of the
manuscript. I also thank
V.~Rivelles for discussions and hospitality at the
department of mathematical physics of the university of S\~ao Paulo (Brazil),
where part of this work was done.

\listrefs
\vfill
\eject
\bye